\documentclass[conference]{IEEEtran}
\usepackage{graphicx}
\usepackage{dblfloatfix}
\usepackage{fixltx2e}
\usepackage{amsmath}

\ifCLASSINFOpdf
\else
\fi
\begin{document}
%
\title{Technique of active phase stabilization for the interferometer with 128 actively selectable paths}

\author{\IEEEauthorblockN{Yu Xu, Jin Lin, Yu-Huai Li, Hui Dai, Sheng-Kai Liao, Cheng-Zhi Peng}
\IEEEauthorblockA{Hefei National Laboratory for Physical Sciences
at the Microscale \\and Department of Modern Physics University of Science and Technology of China
Hefei, 230026, China\\
Chinese Academy of Sciences (CAS) Center for Excellence and Synergetic Innovation Center \\in Quantum Information and Quantum Physics, University of Science and Technology of China, Shanghai 201315, China\\
Email: xuyu0922@mail.ustc.edu.cn}}


%


\maketitle

\begin{abstract}
A variable-delay optical interferometer with 128 actively selectable delays and a technique of active phase stabilization are innovatively designed and applied for the first time in the experiment of round-robin differential phase shift quantum key distribution (RRDPS-QKD). According to the RRDPS protocol, larger number of delay channels in interferometer can ensure higher tolerance of bit errors, eliminating the fundamental threshold of bit error rate of 11\% of traditional BB84 protocol. Considering that the implementation of RRDPS protocol relys on the realizaiton of the interferometer, therefore, an interferometer with 128 selectable delay paths is constructed and demands the ability of fast switching at the rate of 10 $kHz$, which requires dynamic stability of multiple paths. Thus, a specific designed phase stabilization technique with closed real time feedback loop is introduced to guarantee the high visibility of interferometer selections dynamically. The active phase stabilization technique employs a phase modulator (PM) driven by a DAC to adjust the relative phase between the two arms of the interferometer. By monitoring photon counting rates of two Up-Conversion Detectors (UCD) at two output ports of the interferometer, a Field Programmable Gate Array (FPGA) calculates and finds the optimal code value for the DAC, maintaining a high visibility of the interferometer every time a new light path is selected. The visibility of most of the 128 interferometer selections can simultaneously be maintained over 96\% during the QKD, which lays the foundation for the RRDPS experiment.
\end{abstract}



%
\IEEEpeerreviewmaketitle

\section{Introduction}
RRDPS-QKD has advantage over the traditional Bennett-Brassard-1984 (BB84) protocol in term of eliminating the fundamental threshold of bit error rate of 11\% \cite{shor2000simple}. The RRDPS protocol has a better tolerance of bit errors, which makes it easier to accomplish QKD in high noise background \cite{ zhang2016practical}, such as the application of long distance free-space communication. However, the implementation of the RRDPS scheme relies on the realization of a variable-delay Mach-Zehnder interferometer, which is extremely sensitive to the mechanical and acoustic vibrations, temperature drift and other disturbances \cite{xavier2011stable}. These disturbances lead to phase imbalances between two arms of the interferometer. Although a frame of high-damping materials have been employed to envelop the interferometer as a passive protection, we still need a solution to eliminate the residual phase instability caused by the drift of the central wavelength of the laser or other low frequency disturbances. Besides, a suitable phase should be modulated on PM immediately when a new light path is selected. Therefore, a system of active phase control with closed feedback loop is designed and implemented in the RRDPS.

In the RRDPS protocol, Alice prepares and sends a train of $L$ weak coherent state pulses encoded with phase $0$ or $\pi$ by local random numbers. Bob splits the $L$-pulse train into two with a beam splitter. He randomly shifts one of the split pulse trains by $r$ pulses where $1 \leq r \leq L-1$, and obtains a detection on pulse $i$ after interfering two pulse trains. The interference act as an Exclusive OR operation. In the protocol, assume the case where a single photon light source is used, the key rate is given by
\begin{equation}
R = Q[1-h(e_{bit})-h(\frac{v_{th}}{L-1})],
\end{equation}
where $R$ is the final key bit per $L$-pulse train, $e_{bit}$ is the bit error rate, $Q$ is the average number of valid detection per $L$-pulse train, which can be measured in experiment. $h(x)$ is the binary entropy function, and $v_{th}$ is an auxiliary parameter that has an upper bound of $(L-1)/2$. Thus, $R$ is positive correlation to the $L$, i.e., larger $L$ ensures higher tolerance of bit errors. For example, for $L=128, v_{th}=1$, $R$ is positive up to $e_{bit}$ = 0.35 \cite{ sasaki2014practical}.

In this paper, a variable delay optical interferometer with 128 actively selectable delays and a technique of active phase stabilization are innovatively designed and applied for the first time in the experiment of RRDPS-QKD.
\begin{figure*}[t]
\centering
\includegraphics[width=1.0\linewidth]{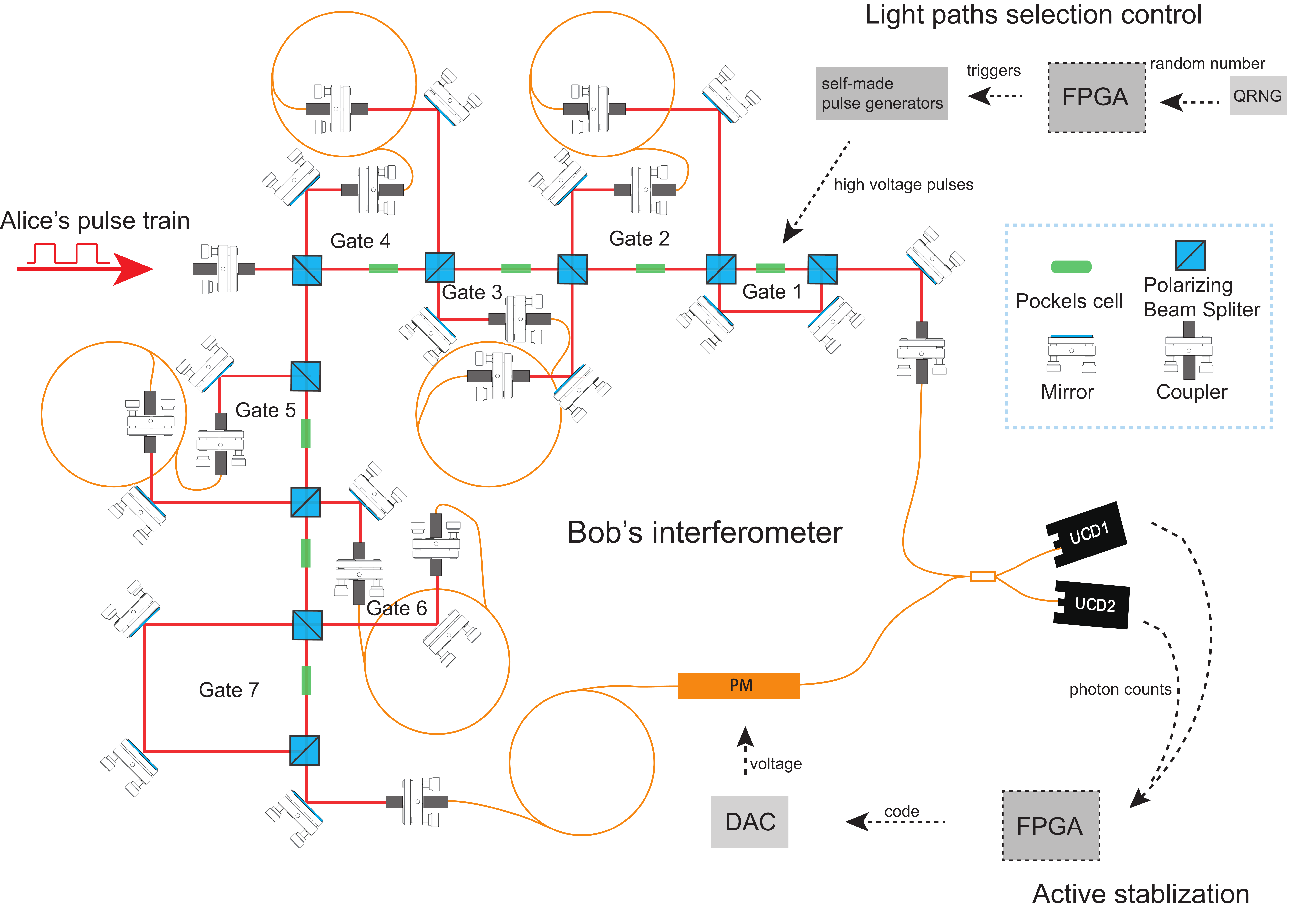}
\caption{The Mach-Zehnder interferometer with Bob's control system.}\label{Fig1}
\end{figure*}

\section{Experiment design}
In our experimental proposal, $L$ is designed to be as big as 128. Therefore, 128 light paths of different lengths should be prepared so as to achieve discrete delay values $r= \{0 ns, 2 ns, 4 ns, 6 ns,\ldots, 254 ns\}$ precisely. Seven delay gates access to seven optical fibers of different lengths are embedded in Bob's interferometer. Each gate switches under the control of a Pockels cell. A Pockels cell consists of two nonlinear crystal in opposite orientation, which makes it behave like a half-wave plate at the half-wave voltage. In order to achieve fast switching between 0 $V$ and half-wave voltage, which is around 2100 $V$ in experiment, seven custom-built RF-MOSFET based high voltage generators are employed to supply the square-shaped 2 $kV$ pulses with repetition rate of 10 $kHz$. These generators are triggered by a FPGA Virtex-6 embedded in the Bob's control system. The triggers' sequence are encoded by a 7-bit random number generated by a physical noise random number generator chip WNG8 on board during QKD process. As shown in Fig.~\ref{Fig1}, thanks to the Bob's control system as well as the high voltage generators, the whole interferometer can transform into any one of the 128 light paths as soon as the random number changes, whose switching rate reaches 10 $kHz$.

\subsection{Hardware design}
\begin{figure}[t]
\centering
\includegraphics[width=1.0\linewidth]{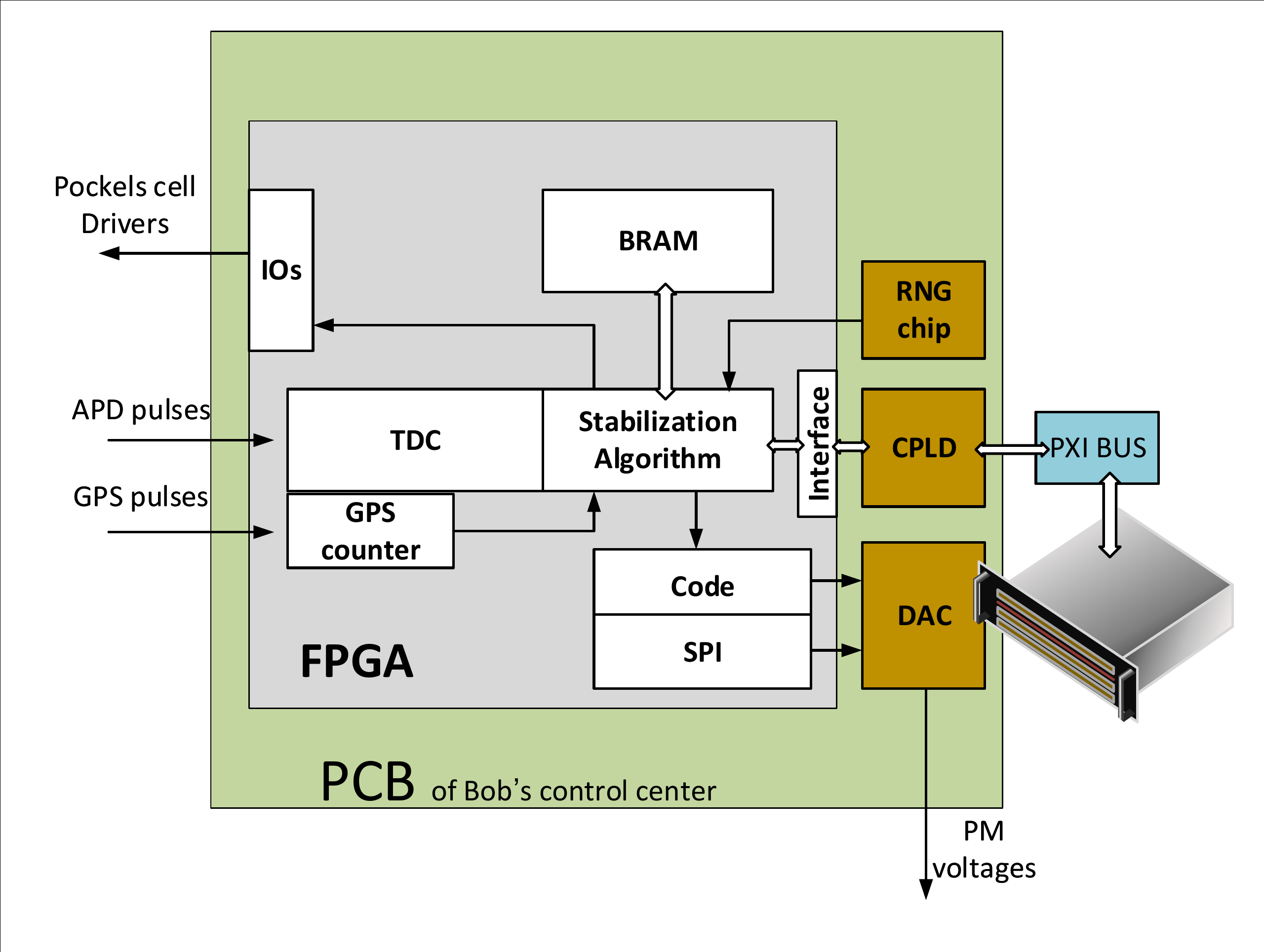}
\caption{The Mach-Zehnder interferometer with Bob's control system.}\label{Fig2}
\end{figure}
To accomplish the RRDPS-QKD experiment, a PCB board is designed to play a role as Bob's control center and embedded in one PXI box. As shown in Fig.~\ref{Fig2}, the control center board is mainly consist of  Virtex-6 FPGA, DAC, CPLD and physical noise random number generator chip (RNG chip). We build several major logic modules in the key component FPGA. The time digital converter (TDC) module is built to distinguish and count the input single photons from APDs, and transmit the counts data to calibration algorithm module. The calibration algorithm module is in charge of calculating input data and manipulating the DAC chip with the help of SPI module and DAC code module. Meanwhile, the algorithm module distributes triggers to 7 Pockels cell drivers corresponding to the random numbers generated by RNG chip and records these random numbers at block ram.
\subsection{Stabilization preparation Stage}
In order to realize real-time phase stabilization, Bob's control system is designed to refresh the optimal data for phase stabilization in first 340 ms of every second, which is named stabilization preparation stage.  The output intensity of one port of the interferometer is defined as $\frac{I}{2}(1+cos(\alpha_{r}))$, where I is the input power and $\alpha_{r}$ is the relative phase between two delay paths. The active phase stabilization is designed to measure $\alpha_{r}$ at stabilization preparation stage and compensate for it by using a PM at the QKD stage. For each delay $r$, an optimal compensation voltage of PM, which is deployed to adjust the relative phase between two arms of the interferometers, should be measured and recorded. And there will be 128 different delays $r$ in total.
\begin{figure}[h]
\centering
\resizebox{9cm}{!}{\includegraphics[scale=1]{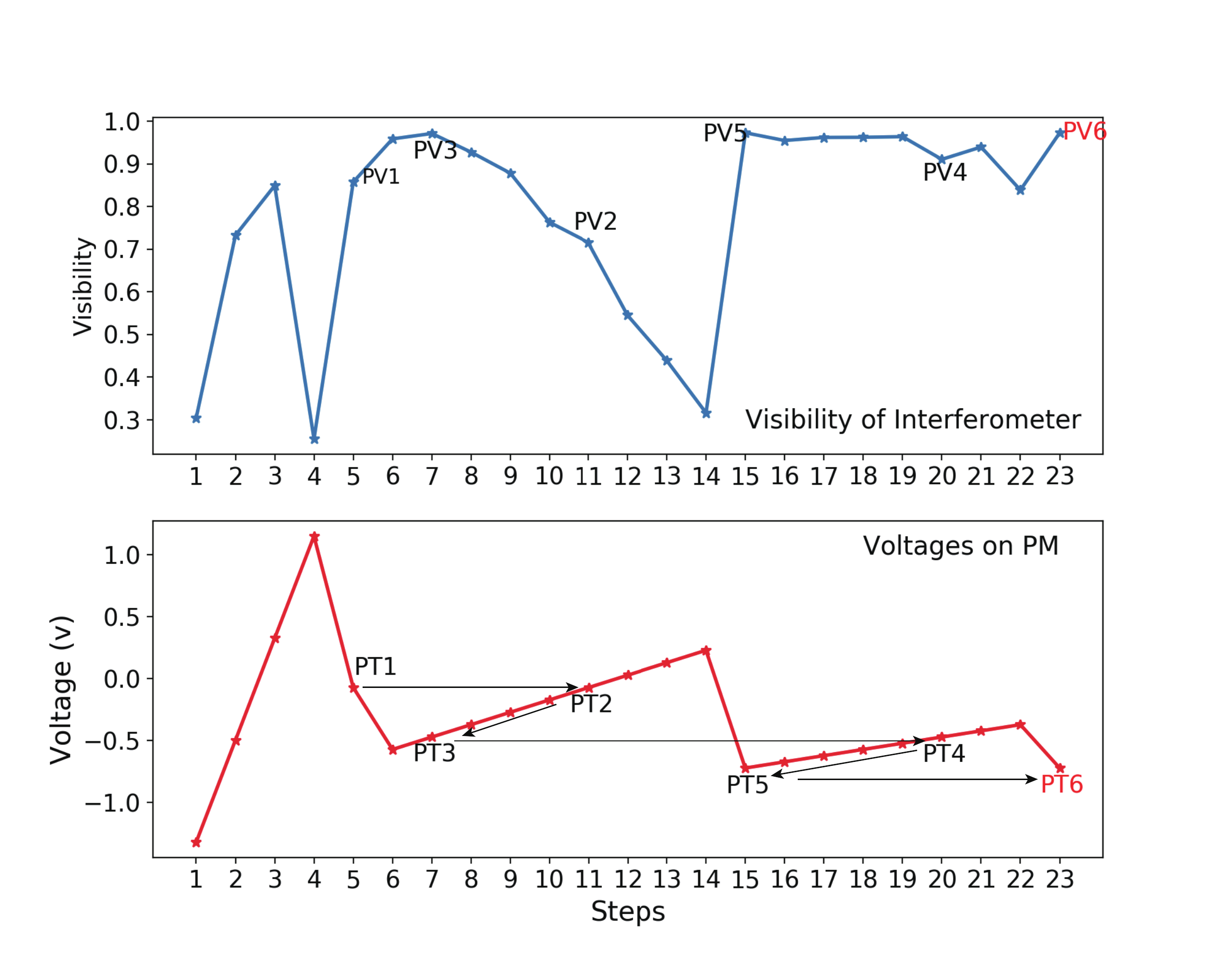}}
\caption{Stabilization algorithm working process. Top subfigure demonstrates the visibility of interferometer when corresponding voltage implemented on PM in bottom subfigure\label{Fig3}}%
\end{figure}
\subsubsection{Stabilization algorithm}
In stabilization preparation stage, Bob's control system activates the specific Pockels cells to go through 128 permutations in order, representing 128 different delays. Each permutation holds for 2.5 ms.  During this period of 2.5 ms, the PM is set to sequentially apply extra phases of $\alpha^{k}_{ext}$. Then using a least-squares method, an initial compensation voltage is obtained in FPGA by figuring out the minimum of the calculation results of
\begin{equation}
   S(\alpha_{r})=\sum_{k}(\frac{I_{k}}{I}-\frac{C^{k}_{1}}{C^{k}_{1}+C^{k}_{2}})^2,
\end{equation}
 where k $\in \{0,1,2,3\}$, representing the four initial step measurements, $I_{k}=\frac{I}{2}(1+cos(\alpha_{r})+cos(\alpha^{k}_{ext}))$ refers the ideal corresponding output intensity of the measurements, $C^{k}_{1}$ and $C^{k}_{2}$ are the photon counts captured by two APDs and recorded by FPGA. Due to imperfections, a final optimal voltage is determined after 2 stage of calibration process after initial calculation. As demonstrated in Fig.~\ref{Fig3}, where the visibility is defined as $\frac{C^{k}_{1}-C^{k}_{2}}{C^{k}_{1}+C^{k}_{2}}$, PT1 is the preliminary result point calculated by the least-squares equation with first four fixed steps (step 1 to 4). And from step 6 to step 14, DAC implements 9 voltages with fixed interval of 0.1 V around PT2 which has the equal voltage value of PT1. This seeking process is named preliminary calibration stage. During preliminary calibration stage, a relatively better working point PT3 is found corresponding to a high visibility PV3. For the reason that voltage value of PT3 sometimes is still not the most optimal working position, 8 more voltages with smaller interval is once more implemented around PT4, which is called secondary calibration stage. Among these 8 precise points, a working points with relatively highest visibility PT5 is figured out. Then, at the last step, PT6, with the same voltage of PT5, is implemented on step 23, in order to double check the result. After these 23 steps changing voltages on PM, an optimal working voltage is determined corresponding to one particular paths interferometer among 128. At the end of the whole stabilization preparation, a reference table is constructed to store 128 new refreshed compensate voltage in this period of one second.

\subsection{QKD Stage}
As soon as the stabilization preparation is finished, Alice and Bob step into the QKD stage. In the remaining 660 $ms/s$ used for QKD, the seven gates turned off or on simultaneously at the rate of 10 $kHz$ to bridge the light paths in accordance with the random number. Meanwhile, every time light paths switch, the PM is immediately set to the corresponding compensation voltages measured in the first 340 $ms$ of this second, ensuring maximum visibility of interferometer. 

\begin{figure}[h]
\centering
\resizebox{8cm}{!}{\includegraphics[scale=1]{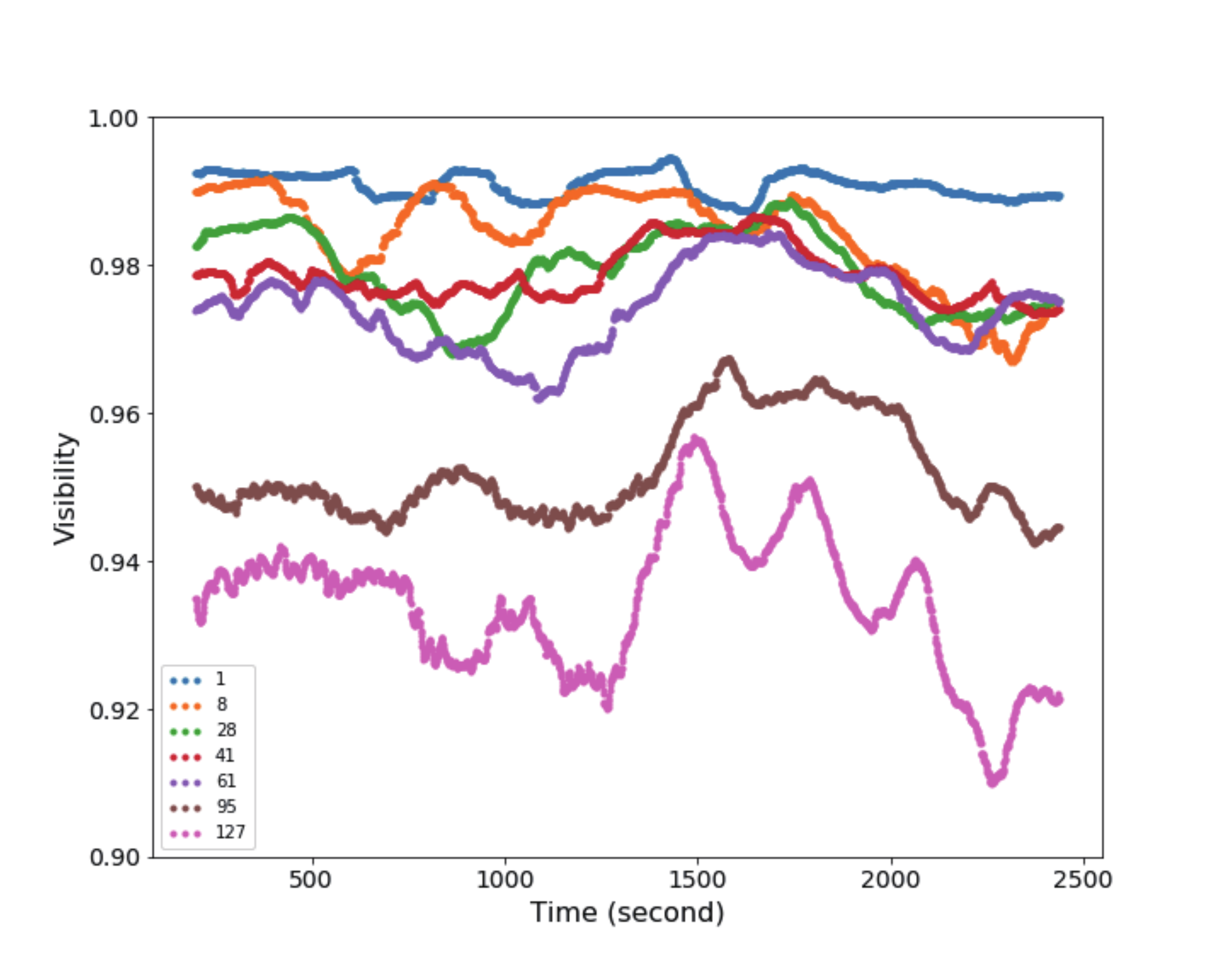}}
\caption{Phase stability with our active phase stabilization technique. Visibilities of interferometer with random delay values $r$ during long time QKD experiment, taking $r$=1,8,28,41,61,95,127 for example.
\label{Fig4}}%
\end{figure}
\section{Conclusion}
As demonstrated in Fig.~\ref{Fig4}, with the help of our active phase stabilization technique, the unequal-arm interferometers with most delay values of $r$, such as $r$ = 1,8, 28, 41, 61, etc can maintain a visibility above 96\% during the QKD stage, in spite of that the visibilities have a weak downward trend as the delay $r$ gets higher. This active phase stabilization technique as well as the actively delays-selectable interferometers design lays the foundation for RRDPS-QKD experiment \cite{li2016experimental} which obtains a final key rate of 15.54 bps with total loss of 18 dB and an error rate of 8.9\%.





\bibliographystyle{IEEEtran}
\bibliography{paper}
%



\end{document}